\documentclass[preprint,superscriptaddress]{revtex4-1}

\usepackage{graphicx}%insertion of eps figures
\usepackage{natbib}%apsrev is used for all the aps journals except rmp
\usepackage[usenames,dvipsnames]{color} % use colored text in latex
\usepackage{amsmath} % Defines extra environments for multiline displayed equations, as well as a number of other enhancements for math
\usepackage{amssymb}% {bm}% bold math
\usepackage{soul}%for underlining which allow text wrappin; also other things
\usepackage{ifthen}%allows defining if then function
\newcommand{\rt}[1]{\sqrt{#1}}

\def\be#1\ee{\begin{equation}#1\end{equation}}
\def\ba#1\ea{\begin{align}#1\end{align}}
\def\bg#1\eg{\begin{gather}#1\end{gather}}
\def\t{\text}

\def\showpersonote{0} %make this flag zero if you want text under \personote to be shown
\newcommand{\personote}[1]{\ifthenelse{\showpersonote=1}{\textcolor{Red}{[[#1]]}}{}}
\def\showaddmat{0} %make this flag zero if you want text under \personote to be shown
\newcommand{\addmat}[1]{\ifthenelse{\showaddmat=1}{\textcolor{Gray}{[[#1]]}}{}}
\def\showinclass{0} %make this flag zero if you want text under \inclass say to be shown
\newcommand{\inclass}[1]{\ifthenelse{\showinclass=1}{\textbf{\textcolor{Brown}{/$\backslash$ #1 /$\backslash$}}}{}}

\begin{document}

\title{Ultrasensitive magnetic field detection using a single artificial atom}

\author{M. Bal}
\affiliation{Institute for Quantum Computing, Department of Physics
and Astronomy, and Waterloo Institute for Nanotechnology, University
of Waterloo, Waterloo, ON, Canada N2L 3G1}

\author{C. Deng}
\affiliation{Institute for Quantum Computing, Department of Physics
and Astronomy, and Waterloo Institute for Nanotechnology, University
of Waterloo, Waterloo, ON, Canada N2L 3G1}

\author{J.-L. Orgiazzi}
\affiliation{Institute for Quantum Computing, Department of Physics
and Astronomy, and Waterloo Institute for Nanotechnology, University
of Waterloo, Waterloo, ON, Canada N2L 3G1}

\author{F. Ong}
\affiliation{Institute for Quantum Computing, Department of Physics
and Astronomy, and Waterloo Institute for Nanotechnology, University
of Waterloo, Waterloo, ON, Canada N2L 3G1}

\author{A. Lupascu \footnotemark[1] \footnotetext[1]{Corresponding author: alupascu@uwaterloo.ca}}
\affiliation{Institute for Quantum Computing, Department of Physics
and Astronomy, and Waterloo Institute for Nanotechnology, University
of Waterloo, Waterloo, ON, Canada N2L 3G1}
%\email{alupascu@uwaterloo.ca}

\date{ May 17, 2012}

\begin{abstract}
%ABSTRACT
Efficient detection of magnetic fields is central to many areas of research and has important practical applications ranging from materials science to geomagnetism. High sensitivity detectors are commonly built using direct current-superconducting quantum interference devices (DC-SQUIDs) or atomic systems. Here we use a single artificial atom to implement an ultrahigh sensitivity magnetometer with a size in the micron range. The artificial atom is a superconducting two-level system at low temperatures, operated in a way similar to atomic magnetometry. The high sensitivity results from quantum coherence combined with strong coupling to magnetic field. By employing projective measurements, we obtain a sensitivity of $2.7\, \t{pT}/\sqrt{\t{Hz}}$ at 10 MHz. We discuss feasible improvements that will increase the sensitivity by over one order of magnitude. The intrinsic sensitivity of this method to AC fields in the 100 kHz - 10 MHz range compares favourably with DC-SQUIDs and atomic magnetometers of equivalent spatial resolution. This result illustrates the potential of artificial quantum systems for sensitive detection and related applications.
\end{abstract}
%\pacs{85.25.Cp%Josephson devices
%, 42.50.Dv %Quantum state engineering and measurements
%, 03.65.Ta%Measurement theory (quantum mechanics)
%, 85.25.Dq%SQUIDs
%%OTHER POSSIBLE OPTIONS
%%03.67.Lx%Quantum computation
%%07.57.Kp%microwave detectors~\cite{taylor_2008_NVDiamondMagn}
%%85.25.Oj%superconducting photodetectors
%}
\maketitle

Sensitive physical measurements are an essential component of modern science and technology. Developments in this area follow closely scientific discovery and provide in
turn tools for practical applications and new research endeavours. Detection of magnetic fields is an area of wide interest, with various applications including medical imaging, geomagnetics, non-destructive materials evaluation, scanning probe microscopy, and electrical measurements~\cite{clarke_2004_1,budker_2007_AtomicMagnReview}. Magnetic field sensors are also enabling tools for fundamental studies of magnetism~\cite{dang_2010_MagntoHighSens}, spin dynamics~\cite{degen_2008_ScanningFieldNV}, and mechanical motion~\cite{arcizet_2011_NVcoupledRes,kolkowitz_2012_CohSensResonatorSpin}.

Sensitive tools to detect magnetic fields are diverse and include direct-current superconducting quantum interference devices (DC-SQUIDs)~\cite{clarke_2004_1}, atomic magnetometers~\cite{budker_2007_AtomicMagnReview}, Hall probes ~\cite{chang_1992_ScanningHall}, and magnetostrictive sensors~\cite{forstner_2012_CavityOptoMagnetometer}. DC-SQUIDs have been established for a long time as very sensitive magnetometers~\cite{clarke_2004_1}. In recent years, significant advances in atomic control led to the development of atomic magnetometers, which presently compete with DC-SQUIDs for magnetic field detection and have the convenience of operation in a room temperature environment~\cite{budker_2007_AtomicMagnReview}. Atomic magnetometers employ ensembles of atoms, with each atom evolving quantum coherently in the field to be measured. A detection method similar to atomic magnetometry can be implemented using nitrogen-vacancy defects (NV centers) in diamond crystals~\cite{taylor_2008_NVDiamondMagn}. Magnetometers based on single NV centers~\cite{maze_2008_NanoMagnSingleSpin,balasubramanian_2008_SingleSpinAmbient} have been demonstrated and shown to have interesting prospects as high spatial resolution high sensitivity detectors. In this paper we demonstrate the use of a \emph{single artificial atom} as an AC magnetometer. The artificial atom, a micron-sized superconducting ring with Josephson junctions, has been studied extensively for quantum computing applications~\cite{clarke_2008_rev-sup-qb}. Our work establishes this system as an ultrasensitive magnetic field detector.

The principle of our approach follows closely magnetometry based on vapour cells and NV centers~\cite{budker_2007_AtomicMagnReview,taylor_2008_NVDiamondMagn}. For a single spin, precession in a field of induction $B$ during time $\tau$ leads to an accumulated phase $\phi=\frac{1}{\hbar}\int_0^\tau m B(t) dt$, with $m$ the magnetic moment. With a coherent control pulse, the spin is rotated so that the projection along the magnetic field depends on $\phi$. A single measurement produces a result $r= \pm 1$, corresponding to the two spin states. For $N$ repetitions, the average value of the measurement $\langle r \rangle=\sin\phi$ and the variance is $1/\rt{N}$. The minimum magnetic field difference, which can be reliably measured, corresponding to a signal to noise ratio of $1$, is $\delta B_{\t{min}}=1/m \tau \rt{N}$. This can be expressed as $\delta B_{\t{min}} = \frac{1}{m}\frac{1}{\rt{\tau T}}\rt{\frac{T_{\t{rep}}}{\tau}} $, with $T_{\t{rep}}$ the repetition time of the state preparation, precession, and measurement procedure described above and $T=N T_{\t{rep}}$ the total measurement time. The sensitivity can thus be expressed as the quantity $\delta B_{\t{min}}\rt{T}$, which has units of $\t{Tesla}/\rt{\t{Hz}}$. The sensitivity increases with the time $\tau$ as long as the evolution is fully quantum coherent; with decoherence taken into account, the optimum is reached when $\tau$ is of the order of the coherence time.

%INSERTING FIG 1
\begin{figure*}[!]
\includegraphics[width=6.5in]{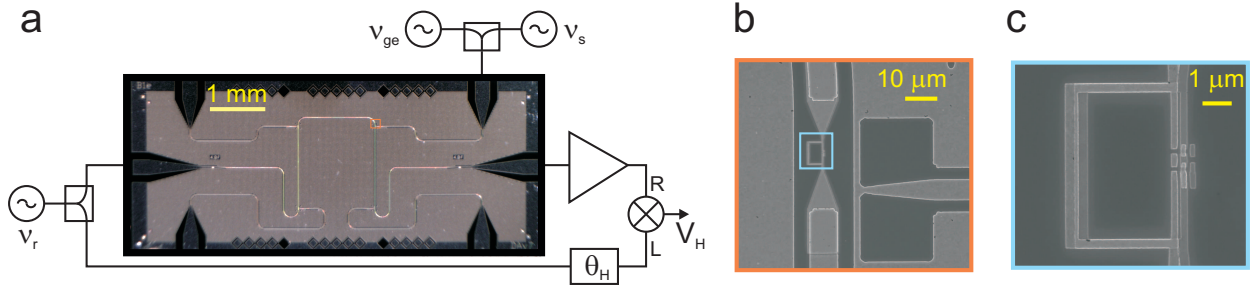}
\caption{\label{fig1} \textbf{Persistent current qubit and control/readout circuit.}\textbf{(a)} Optical microscope picture of a microfabricated device nominally identical to the device used in this work and schematic representation of the control and readout circuit. The qubit is at the position indicated by the orange rectangle. A coplanar waveguide resonator is used for qubit readout: microwaves at frequency $\nu_r$ are applied at the input port (left) and the transmitted wave at the output port (right) is amplified and down converted to determine the homodyne voltage $V_H$. The detected quadrature can be adjusted by changing the phase $\theta_H$. A coplanar waveguide (top right port) is used to send microwave signals at frequency $\nu_{\t{ge}}$ for coherent control of the qubit as well as the AC voltage at frequency $\nu_s$ used to induce the measured magnetic field. \textbf{(b)} Scanning electron microscope picture showing a zoom in the region of the PCQ, corresponding to the orange rectangle in (a). The qubit is the rectangular loop at the center, coupled to the center line of the coplanar waveguide resonator. On the right, the termination of the line for qubit control and AC field (top right port in (a)) is shown. \textbf{(c)} Zoom of the image in panel b in the region indicated by the blue rectangle, showing the PCQ.}
\end{figure*}

In magnetometers based on this principle, the two spin states of interest are hyperfine or Zeeman levels in alkali~\cite{budker_2007_AtomicMagnReview} or ground states in NV centers in diamond~\cite{taylor_2008_NVDiamondMagn}. Detection is done most commonly using an ensemble of atoms/defects manipulated and detected independently. With $N_{\t{at}}$ atoms, sensitivity is further enhanced by a factor $1/\rt{N_{\t{at}}}= 1/\rt {V n}$, with $V$ the volume and $n$ the density of the cloud. The $1/\rt {V}$ dependence of sensitivity implies that there is a tradeoff between sensitivity and spatial resolution. The same tradeoff also appears in magnetic field sensors based on DC SQUIDs~\cite{clarke_2004_1}.

Here we use a single artificial atom to implement an ultrasensitive magnetometer for AC magnetic fields. The artificial atom is a persistent current quantum bit (PCQ)~\cite{mooij_1999_1}, operated in this experiment at a temperature of $43\,\t{mK}$. This type of system has been under intense investigation for applications in quantum information processing~\cite{clarke_2008_rev-sup-qb}. The PCQ is a superconducting quantum ring, with typical size in the micron range, interrupted by three Josephson junctions (see Fig.1c). The two lowest energy eigenstates of the PCQ are characterized by a persistent current $I_p$, flowing either anticlockwise or clockwise in the qubit ring. In the basis of the persistent current states, the qubit Hamiltonian is given by $H_{qb}=-\frac{h \Delta}{2}\sigma_x-I_p (\Phi - \Phi_0/2)\sigma_z$ where $\Phi$ is the magnetic flux applied to the qubit ring, $\Phi_0=h/2e$ is the flux quantum, and $h \Delta$ is the minimum energy level splitting of the qubit, which occurs at the symmetry point $\Phi=\Phi_0/2$ (see Fig.3a). The effective magnetic moment of the qubit is $m=\left|\frac{dE_{ge}}{dB}\right|$ with $E_{ge}$ the energy-level difference between the excited (e) and ground (g) states and $B=\Phi/A_{\t{qb}}$ the magnetic field applied to the qubit ring. The magnetic moment $m$ is given by $m=\frac{\rt{\nu_{ge}^2-\Delta^2}}{\nu_{ge}}2 I_p A_{\t{qb}}$, with $\nu_{ge}=E_{ge}/h$ the qubit transition frequency. For our qubit, characterized by $I_p=139\,\t{nA}$, $A_{\t{qb}}=24.5\:\mu\t{m}^2$, and $\Delta=10.11\,\t{GHz}$, and operated at $\nu_\t{ge}=11.24\,\t{GHz}$, $m$ reaches the value $3.2\times 10^5\:\mu_B$. This large magnetic moment enables a large sensitivity, despite the coherence time of the PCQ being shorter than in typical atomic systems~\cite{budker_2007_AtomicMagnReview,taylor_2008_NVDiamondMagn}.

To coherently control the quantum state of the PCQ, microwave fields at the transition frequency $\nu_{ge}$ are applied through an on-chip waveguide terminated in a low inductance line (see Fig.\ref{fig1}a and b). In a frame rotating at the qubit transition frequency, the microwave field acts as a fictitious magnetic field that induces rotation of the qubit around an axis in the $xy$ plane; the orientation of the rotation axis in this plane depends on the phase of the driving field. In the same frame, a change in the magnetic field $B$ applied perpendicularly to the qubit loop results in a fictitious magnetic field along the $z$ axis. Below, we use $\theta_{\overrightarrow{n}}$ to denote a rotation of angle $\theta$ around an axis defined by the vector $\overrightarrow{n}$.

Quantum measurement of the PCQ is done by using a circuit-quantum electrodynamics setup~\cite{wallraff_2004_1,abdumalikov_2009_flux-qubit,niemczyk_2010_cqedstrong}. The qubit is inductively coupled to a superconducting resonator, with a resonance frequency $\nu_\t{res}=6.602\,\t{GHz}$, significantly lower than the qubit transition frequency $\nu_\t{ge}$, and quality factor $Q=4,000$. A microwave readout pulse of duration $T_r$ and frequency $\nu_\t{r}=\nu_{\t{res}}$ is sent to the qubit. The complex amplitude of the transmitted pulse, as determined in  a homodyne measurement~\cite{wallraff_2005_1}, is averaged over the duration $T_r$ of the readout pulse. In Fig.~2 we present the results of the qubit measurement. We only show one quadrature of the transmitted voltage, $V_H$; the axis for this quadrature is chosen such that it optimizes the measured signal (see Fig.~1a). The qubit is prepared either in the ground state by allowing for a waiting time much longer than the energy relaxation time $T_1=1.2\:\mu\t{s}$ (Fig 2a) or in the excited state by including a $\pi_x$ pulse (Fig 2b). The distribution of the values of the homodyne voltage $V_H$ for $10^4$ repetitions of the sequence is shown in Fig. 2c and 2d respectively. The distribution is bimodal, with the two modes corresponding to the qubit energy eigenstates. A threshold can be used to separate the distribution into a part labeled $r=-1$ and the complementary part labeled $r=1$. The threshold is chosen so that it optimizes the readout contrast, which is the difference of the conditional probabilities $P(r=-1|\:\t{e})-P(r=-1|\:\t{g})$. The maximum contrast is $62\,\%$. This high readout fidelity is essential for the sensitivity of the detector.

%INSERTING FIG 2
\begin{figure}[!]
\includegraphics[width=3.4in]{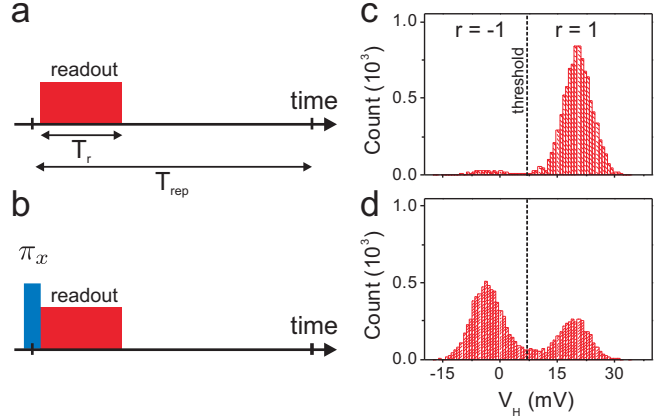}
\caption{\label{fig2} \textbf{Characterization of readout efficiency for the persistent current qubit.} \textbf{(a,b)} Protocol used for characterization of readout for the ground state (a) and the excited state (b). The excited state is prepared by applying a $\pi$ pulse prior to qubit readout. The readout time is $T_r=400\,\t{ns}$ and the repetition time $T_{\t{rep}}=20\,\mu\t{s}$ is chosen significantly longer than the qubit relaxation time $T_1$. \textbf{(c,d)} Histograms of the homodyne voltage values $V_H$ for preparation in the ground state (c) and the excited state (d). The horizontal dashed line indicates the position of the threshold used to separate values labeled $r=1$ and $r=-1$ which are associated with the ground and excited states of the qubit respectively. }
\end{figure}

In typical magnetometers, as introduced above, free precession in a magnetic field is employed. This procedure is adapted to detection of low-frequency fields, ranging from DC to the inverse of the repetition time, $T_{\t{rep}}^{-1}$. The sensitivity is proportional to $\frac{\rt{T_\t{rep}}}{T_2^*}$, where $T_2^*$ is the Ramsey coherence time~\cite{yoshihara_2006_1}. This coherence time is short in the PCQ due to the presence of low-frequency magnetic flux noise~\cite{yoshihara_2006_1,bylander_2011_noisePCQ}. Low-frequency noise sets the ultimate limit for measurement of low frequency fields, a situation also encountered for DC-SQUIDs~\cite{clarke_2004_1}. For this reason, we focus here on detection of AC magnetic fields. The procedure is illustrated in Fig. 4a. A $ \left( \frac{\pi}{2}\right) _x - \pi_x - \left( \frac{\pi}{2}\right) _y $ sequence of pulses (also called a spin-echo sequence) is applied to the qubit at times $0$, $\tau/2$, and $\tau$ respectively . The qubit phase precession is given by $\phi=\pi-\frac{1}{\hbar}\int_0^{\frac{\tau}{2}}m B(t) dt+\frac{1}{\hbar}\int_{\frac{\tau}{2}}^{\tau}m B(t) dt$. The acquired  phase is optimized when the frequency $\nu_\t{s}$ of the detected field is equal to $\tau^{-1}$. The coherence time during the spin-echo sequence, $T_{2}$, is significantly longer than the Ramsey time $T_2^*$~\cite{yoshihara_2006_1}, which renders the sensitivity to AC field higher than for low frequency fields.

We performed measurements of the evolution of the qubit with a spin-echo sequence of varying total time $\tau$ with an AC magnetic field of frequency $\nu_\t{s}=\tau^{-1}$ applied to the qubit in phase with the spin-echo sequence, as shown in Fig. 4a. The AC magnetic field is applied through the same control line as used for qubit excitation (see Fig. 1a), by applying a voltage of amplitude $V_{\t{ac}}$. The average value of the homodyne voltage $V_H$ is shown in Fig. 3b as a function of the spin-echo sequence time and the voltage amplitude $V_{\t{ac}}$. We observe oscillations as a function of the time $\tau$, with a frequency $\delta \nu_{ge}$, which is proportional to the amplitude $V_{\t{ac}}$ (see Fig. 3c).

%INSERTING FIG 3
\begin{figure*}[!]
\includegraphics[width=6.0in]{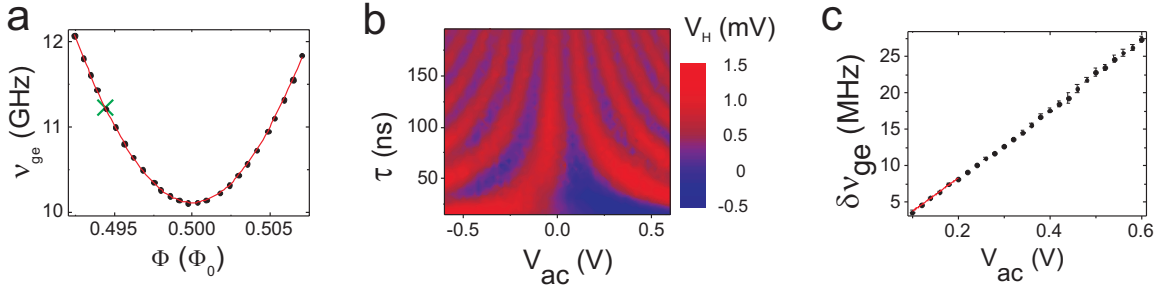}
\caption{\label{fig3} \textbf{Measurements of qubit spectroscopy and spin-echo precession in an AC magnetic field.} \textbf{(a)} The qubit transition frequency $\nu_{ge}$ versus the magnetic flux $\Phi$ applied to the qubit ring. The continuous line is a fit with the expression $\nu_{ge}=\rt{\Delta^2+\left( \frac{2I_p}{h}\left(\Phi-\frac{\Phi_0}{2}\right) \right)^2}$, which allows extracting the qubit parameters $\Delta$ and $I_p$ given in the text. The cross indicates the position, at $\nu_{\t{ge}}=11.24\,\t{GHz}$, where the AC magnetic field measurements are performed. \textbf{(b)} Measurement of the qubit evolution under a spin-echo sequence versus the total duration $\tau$ and the amplitude $V_{\t{ac}}$ of an AC field synchronous with the spin-echo pulses. The measurement sequence is as shown in Fig. 4a, with $T_{\t{rep}}=10\,\mu\t{s}$ and $T_{\t{r}}=1\,\mu \t{s}$. The homodyne voltage $V_H$ is averaged over $N=10^{4}$ repetitions. \textbf{(c)} Plot of the spin-echo oscillation frequency $\delta \nu_{\t{ge}}$ versus the AC field amplitude. The oscillation frequency is obtained by fitting the data in panel (b) with a damped sine; the error bars represent fit parameter errors. The continuous line shows a linear fit in the region of small AC amplitude, relevant for small signal detection, yielding the conversion factor $\frac{\partial \nu_{\t{ge}}}{\partial V_{\t{ac}}}=46.1\, \t{kHz/mV}$. At large value of $V_{\t{ac}}$ the slope changes slightly due to the influence the AC field has on the applied pulses. }
\end{figure*}

We next proceed to the characterization of magnetic field detection using the protocol illustrated in Fig. 4a. The output of the detector is the binary signal $r$. The noise in $r$ reflects the stochastic nature of quantum measurement; in atomic magnetometry this noise is termed projection noise ~\cite{budker_2007_AtomicMagnReview}. In the following we express the results of detection in terms of the magnetic flux $\Phi$ applied to the qubit, as this facilitates the comparison with DC-SQUIDs. The noise in $r$ results in an equivalent noise in $\Phi$ characterized by the spectral density $S_{\Phi}=S_r/\left(\frac{\partial \langle r\rangle}{\partial \Phi}\right)^2$. Here $S_r$ is the single-sided spectral density of the noise in the qubit readout and $\frac{\partial \langle r\rangle}{\partial \Phi}$ is the transfer function of the detector, with $\langle r \rangle$ the average value of $r$. The transfer function can be expressed as
$\frac{\partial \langle r\rangle}{\partial \Phi} = \frac{\partial \langle r\rangle}{\partial V_{\t{ac}}} \frac{\partial \nu_{\t{ge}}}{\partial \Phi} \left( \frac{\partial \nu_{ge}}{\partial V_{\t{ac}}} \right)^{-1}$ where $\frac{\partial \nu_{ge}}{\partial V_{\t{ac}}}$ is determined from the spin-echo measurements shown in Fig. 3c and $\frac{\partial \nu_{ge}}{\partial \Phi}$ is determined from qubit spectroscopy (see Fig. 3a). The conversion factor $\frac{\partial \langle r\rangle}{\partial V_{\t{ac}}}$ is determined from the measurements shown in Fig.~4b. In this way we determine the equivalent detector input noise fully from a set of measurable quantities, without any assumption on coupling of the field to the qubit. We use a spin-echo control sequence time $\tau=121\,\t{ns}$  chosen to correspond approximately to the optimal value for the measured qubit coherence time. We note that in this experiment sample coherence was affected by a two-level fluctuator, as shown by the structure of the spectroscopy peak and also by the fact that there are discrepancies between the observed spin-echo decay and the expected Gaussian law~\cite{yoshihara_2006_1}. We use nevertheless the dependence $e^{-\tau/(2T_1)}e^{-\tau^2/T_2^2}$ to fit the envelope of the spin-echo oscillations and extract $T_2=157 \pm 6\, \t{ns}$. Figure 4c shows $\rt{S_r}$, as obtained by taking the power spectral density of the $r$ vs time signal, and the calculated equivalent input detector noise $S_\Phi^{1/2}$. A theoretical calculation of the sensitivity taking into account the finite measurement fidelity and the experimentally characterized decoherence during the spin echo sequence predicts a flat noise spectrum with a value of $5.2\times 10^{-8}\,\Phi_0/\rt{Hz}$, in good agreement with the experimentally measured value of $8\times 10^{-8}\,\Phi_0/\rt{Hz}$ at high frequency. In the low-frequency region, excess noise is observed due to electronics drifts and interference. For fast detection, the equivalent flux noise is largely dominated by the value in the high frequency limit. For magnetic field detection, the sensitivity can be expressed as $S_B^{1/2}=S_\Phi^{1/2}/A_{\t{qb}}$, yielding $S_B^{1/2}=6.8\, \t{pT}/\sqrt{\t{Hz}}$.

The detection sensitivity obtained above is partly limited by the large ratio $T_{\t{rep}}/\tau$, as imposed by $T_{\t{rep}}\gg T_1$, required to initialize the qubit by energy relaxation. To reduce the overhead time, we introduce another measurement scheme in which we use the correlator $c_i=r_{i+1} r_{i}$ as the detector output, with $r_{i}$ and $r_{i+1}$ two consecutive measurement results at steps $i$ and $i+1$. This is motivated by the fact that an ideal projective readout prepares the qubit in an energy eigenstate. The measurement result $r_{i}$ is random. However, the product $r_i r_{i+1}$ only depends on qubit evolution if energy relaxation is neglected. Our measurement has a limited efficiency and the projection fidelity is lower for $r=-1$ due to energy relaxation during measurements~\cite{lupascu_2007_1,picot_2008_rel-meas-qb}. The repetition time of the sequence $T_{\t{rep}}$ is experimentally optimized to balance two competing effects: a long $T_{\t{rep}}$ results in additional qubit relaxation, which reduces qubit projection when $r=-1$; a short $T_{\t{rep}}$ leads to additional decoherence of the qubit, presumably due to photon number fluctuations in the resonator~\cite{schuster_2005_1}. We obtain an optimum detection efficiency for $T_{\t{rep}}=1000\,\t{ns}$ and $\tau=100\,\t{ns}$. We calculate the magnetic field noise referred to detector output using $S_{\Phi}=S_c/\left(  \frac{\partial \langle c\rangle}{\partial V_{\t{ac}}} \frac{\partial \nu_{ge}}{\partial \Phi} \left( \frac{ \partial \nu_{ge}}{\partial V_{\t{ac}}} \right)^{-1}  \right)^2$, similar to the scheme based on qubit reset using energy relaxation. The conversion factor $\frac{\partial \langle c\rangle}{\partial V_{\t{ac}}}$ is determined from the measurement shown in Fig. 4e. The magnetic field noise is shown in Fig. 4f. A significant improvement of sensitivity is achieved compared to the result obtained using qubit reset  by relaxation, as shown in Fig 4c. The equivalent flux noise, averaged over the full frequency interval, is $3.3\times 10^{-8}\,\Phi_0/\rt{Hz}$. Excess noise is observed at low frequency as well, however the magnitude is significantly lower than for the data shown in Fig.~4c, due to the fact that low frequency fluctuations in the detection system are removed by using the correlations. The magnetic field detection sensitivity reaches $S_B^{1/2}=2.7\, \t{pT}/\sqrt{\t{Hz}}$. The improvement in detection efficiency is due primarily to the reduction in duty cycle.

%INSERTING FIG 4
\begin{figure*}[!]
\includegraphics[width=6.8in]{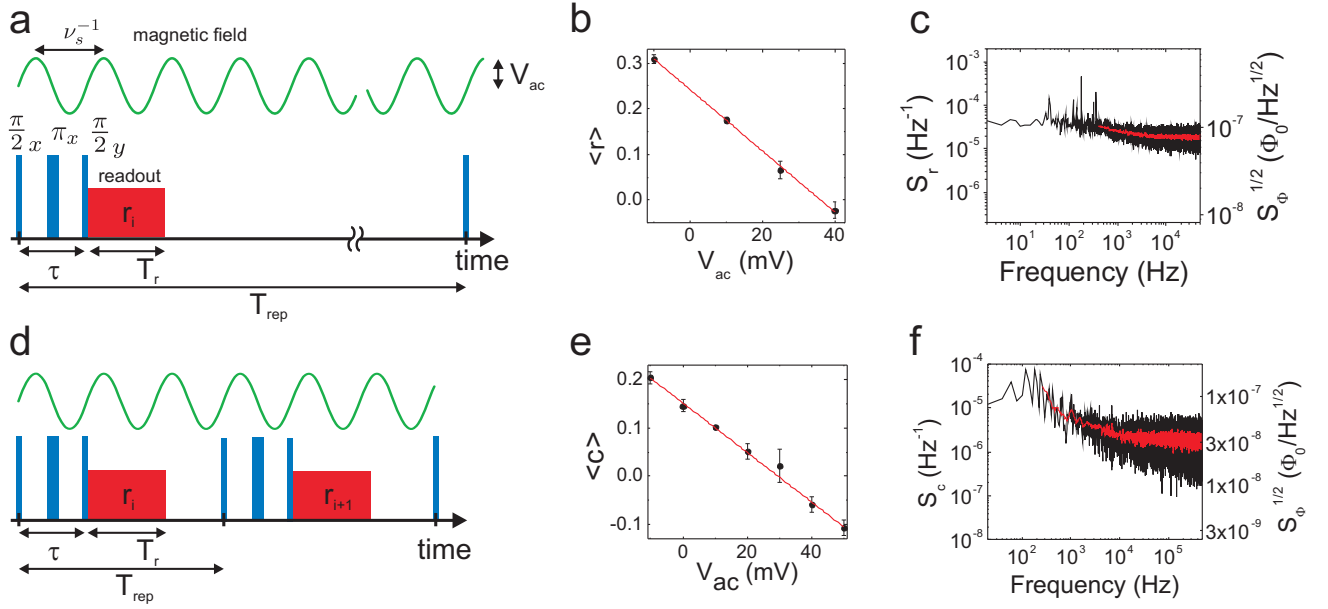}
\caption{\label{fig4} \textbf{Measurement of magnetic flux detection sensitivity.} \textbf{(a)} Qubit control and measurement sequence for the scheme with qubit reset based on energy relaxation. The repetition time $T_{\t{rep}}=10\,\mu\t{s}$ is chosen large enough so that the qubit is prepared in the ground state prior to the control pulse. The qubit is controlled with a spin-echo pulse sequence of duration $\tau=121\,\t{ns}$. A magnetic field, of amplitude proportional to the AC voltage amplitude $V_{ac}$, is applied synchronously to the spin-echo pulse. The measurement result $r_i=\pm 1$ is recorded at each repetition. \textbf{(b,c)} Results from 20 series of measurements, each containing $5\times 10^5$ repetitions of the sequence shown in (a). Panel (b) shows the average $\langle r \rangle$ over the 20 series; the error bars are the standard deviation of single-series average results. The black line in  panel (c) shows the spectral density of the noise of the detector output, $S_r$, determined by averaging results over the $20$ series. The red line is an adjacent points average over a 50 Hz window. The right axis indicates the equivalent value of the detector input noise $S_{\Phi}$. \textbf{(d)} Qubit control and measurement sequence for the detection scheme based on measurement of correlations. The repetition time $T_{\t{rep}}=1\,\mu\t{s}$ and the spin-echo time is $\tau=100\,\t{ns}$. The relevant signal is the product $c_i=r_{i+1}r_i$ of two consecutive measurement results. \textbf{(e,f)} Results from 4 series of measurements, each containing $50,000$ repetitions of the sequence shown in (d). Panel (e) shows the average $\langle c \rangle$ over the 4 series; the error bars are the standard deviation of single-series averages. The black line in  panel (c) shows the spectral density of the noise of the detector output, $S_c$, determined by averaging results over the $4$ series. The red line is an adjacent points average over a 500 Hz window. The right axis indicates the equivalent value of the detector input noise $S_{\Phi}$.}
\end{figure*}

These results establish the PCQ as an ultrasensitive AC magnetic field detector. In the following a discussion is given of how our detector compares with other types of sensors. We focus on the comparison with DC-SQUIDs and atom-based detectors. For DC-SQUIDs, the proper figure of merit related to magnetometry applications is the energy sensitivity $S_{\Phi}/2L$ where $L$ is the loop inductance of the DC-SQUID~\cite{clarke_2004_1}. This same figure of merit is adapted to the PCQ, since similar magnetic field coupling methods can be used. In DC-SQUIDs, $S_\Phi$ is dominated at low frequency by flux noise and reaches a constant value in the high frequency region, which is taken as the relevant figure for noise. By energy sensitivity, our detector, with a loop inductance $L=27\,\t{pH}$, compares favorably with the DC-SQUIDs in ~\cite{awschalom_2008_LowNoiseSQUID} and ~\cite{wellstood_1989_HotElectronSquids} operated at temperatures of $290\,\t{mK}$ and $25\,\t{mK}$ respectively. We also note that optimization of the energy sensitivity in our case can in principle be done by increasing the loop inductance, as flux noise was observed in general not to scale up with loop size, an aspect favourable for magnetometry~\cite{clarke_2004_1}.

For a comparison with atomic magnetometers and NV center based detectors the most adapted figure of merit is the quantity $\delta B_{\t{min}}\rt{T}\rt{V}$, which combines the sensitivity $\delta B_{\t{min}}\rt{T}$ and the detector volume $V$~\cite{budker_2007_AtomicMagnReview}. Atomic magnetometers based on vapour cells have very high sensitivity, achieved usually with volumes of the order of $1\,\t{cm}^3$, when numbers of the order of $0.1-1\,\t{fT}/\rt{\t{Hz}}$~\cite{dang_2010_MagntoHighSens,kominis_2003_SubfemtoteslaMagn} are reached. When extrapolated to volumes of $\approx 1\,\mu\t{m}^3$, corresponding to the PCQ detector used in this work, the ultimate theoretical limit to sensitivity is of the order of $1\,\t{pT}/\rt{\t{Hz}}$~\cite{shah_2007_SubPicoTeslaMagn}. More recently, magnetic field detection based on cold atoms has been explored as well~\cite{wildermuth_2005_BecMagnetometry,vengalattore_2007_HRMagnBEC}. Using a Bose-Einstein condensate (BEC), a sensitivity of $8.3\,\t{pT}/\rt{\t{Hz}}$ for a measurement area of $120\,\mu\t{m}$ was obtained in ~\cite{wildermuth_2005_BecMagnetometry}. The fundamental limit for sensing using a BEC with a resolution of  a few micrometers~\cite{wildermuth_2006_BECElectricAndMagnSensing} is in the $\t{pT}/\rt{\t{Hz}}$ range. We note that atomic magnetometers operate typically at low frequency, below 1~kHz; methods exist to extend the operation frequency to hundreds of kHz~\cite{lee_2006_subfemtoRFMagn}. NV centers in diamond~\cite{taylor_2008_NVDiamondMagn} have recently emerged as an ultrasensitive method for magnetometry. They combine the advantage of the possibility to work at room temperature and a spatial resolution that can be changed from the nanometer range (for single NV center operation)~\cite{maze_2008_NanoMagnSingleSpin,balasubramanian_2008_SingleSpinAmbient,deLange_2011_SingleSpinSensingMultipulse} up to large scale by using a spin ensemble. Decoherence due to paramagnetic impurities limits the flux detection efficiency to $0.250\,\t{fT}/\rt{\t{Hz}}\,\t{cm}^{3/2}$, optimal for AC fields at frequencies of the order of 100~kHz~\cite{taylor_2008_NVDiamondMagn}, which results in a sensitivity of $10\t{pT}/\rt{Hz}$ for a detection volume of the order of $1\,\mu \t{m}^3$.

%INSERTING FIG 5
\begin{figure}[!]
\includegraphics[width=3.1in]{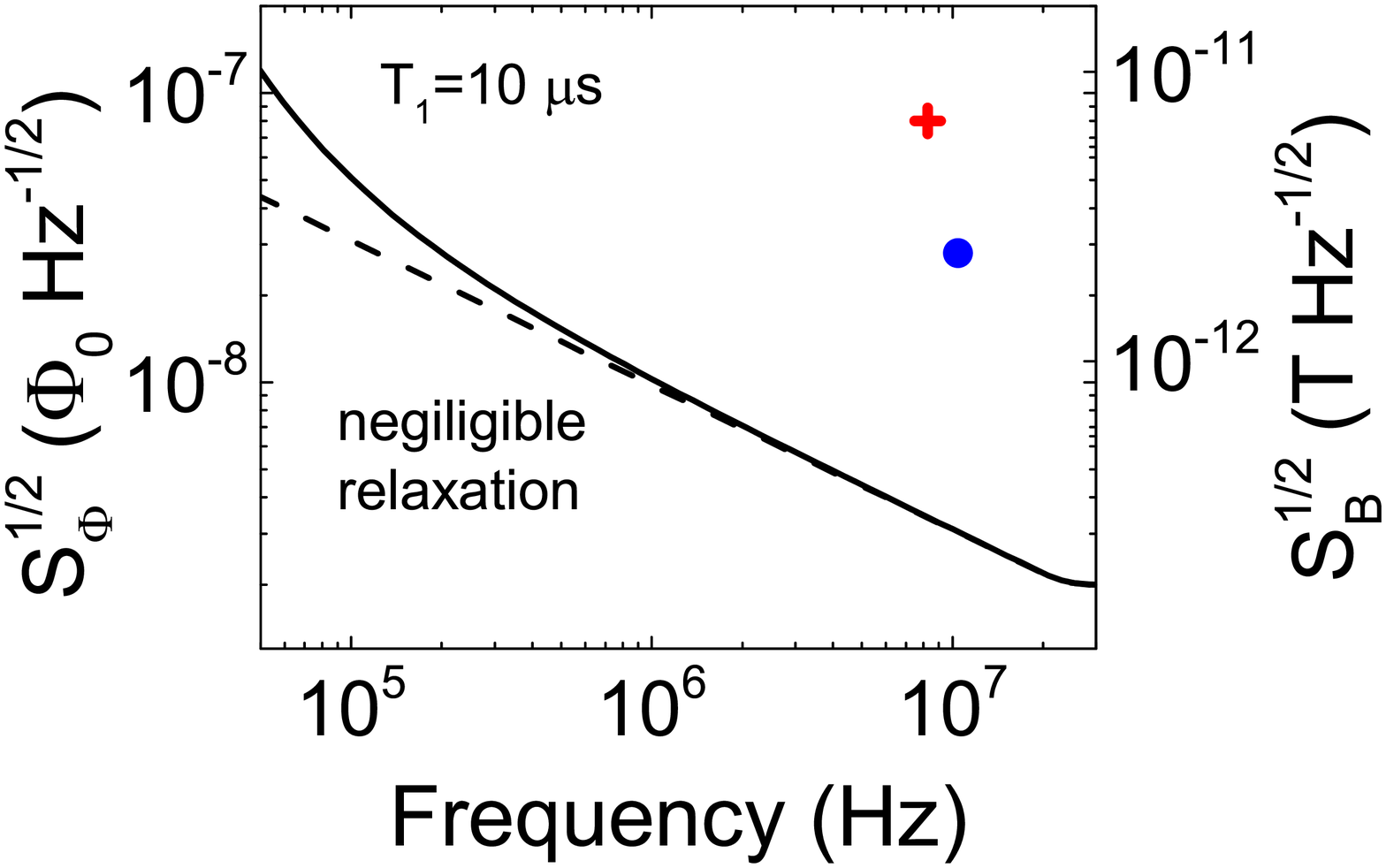}
\caption{\label{fig4} \textbf{Magnetic flux and field detection sensitivity - present results and ideal case.} The calculated sensitivity
for a PCQ for the case of ideal measurements and with a flux noise of $3.8\,\mu \Phi_0/\rt{\t{Hz}}$ is given for the case of a relaxation
time $T_1=10\,\mu\t{s}$ (continuous line) and for the case of negligible relaxation time (dashed line). The experimental results reported
here are indicated by the red cross (qubit reset done by energy relaxation) and blue dot (correlation measurements).}
\end{figure}

Detection of magnetic fields using the PCQ was demonstrated here for AC magnetic fields at $\approx 10\, \t{MHz}$. Detection over a wide range of frequencies is possible. At a given frequency $\nu_\t{s}$, optimizing the detection sensitivity requires tuning of the magnetic moment of the qubit. This can be achieved in situ by changing the qubit transition frequency $\nu_{\t{ge}}$. The optimal sensitivity for a PCQ is calculated here with a set of realistic assumptions on parameters and feasible improvements of control and decoherence. Firstly, the persistent current $I_p$ and minimum energy level splitting $\Delta$ are taken to have values as the PCQ in this experiment. Secondly, ideal projective measurements are used; projective measurements nearly reaching perfect fidelity have been demonstrated for superconducting qubits (see \emph{eg} ~\cite{lupascu_2007_1}). Thirdly, the duty cycle ($\tau/T_{\t{rep}}$) is set equal to one. This can be achieved by shortening readout times, as enabled by nearly quantum limited amplifiers~\cite{castellanosbeltran_2008_ParAmp,bergeal_2010_PhasePresNearQLimit}, and by replacing spin-echo sequence with more complex control pulse schemes~\cite{deLange_2011_SingleSpinSensingMultipulse,bylander_2011_noisePCQ,taylor_2008_NVDiamondMagn}. Finally, qubit pure dephasing is assumed to be limited by 1/f flux noise (see ~\cite{yoshihara_2006_1}), with a spectral density given by $\left(3.8\,\mu\Phi_0\right)^2/f[ \t{Hz}]$, which is the value that would explain the observed spin echo decay time at $\nu_\t{ge}=11.24\,\t{GHz}$ in our experiment if flux noise was the only noise contribution. This level of flux noise is larger than measured values in smaller area superconducting rings~\cite{wellstood_1987_dcsquidnoise,yoshihara_2006_1,bylander_2011_noisePCQ}. It is very likely that in our experiment the flux noise is significantly lower than this upper bound and that charge noise plays a major role, due to the low Josephson energy in this device. Straightforward changes in design will allow reducing the influence of charge noise to negligible levels. With these assumptions taken into account, the calculated optimal sensitivity is plotted in Figure~5 for a PCQ with an energy relaxation time of $10\,\mu \t{s}$ (attained in ~\cite{bylander_2011_noisePCQ}) and for the case where relaxation is neglected. This calculation shows that with respect to the experimental results reported here, more than one order of magnitude of improvement is possible by feasible changes of the experimental setup. The detection efficiency decreases with frequency, due to the ultimate limit imposed by 1/f noise. Nevertheless, this detector has a very high intrinsic sensitivity for measurements in the range from tens of kHz to tens of MHz. Possible future developments on increase of coherence times of superconducting qubits will increase the efficiency and the useful frequency range even further.

In conclusion, we demonstrated a high sensitivity magnetometer based on an artificial atom. The high magnetic field sensitivity combined with the micron spatial resolution is relevant to applications such as detection of electron spin resonance, scanning probe microscopy, and sensitive current and voltage amplifiers~\cite{clarke_2004_1}. This detector is particularly interesting for exploring the dynamics of quantum systems at low temperatures with minimal backaction. The results here illustrate the potential that artificial quantum systems have for quantum sensing.

\section{Methods}
The PCQ presented in this work is realized using a two-step fabrication process. In the first step, a resist layer is applied on a silicon wafer and patterned using optical lithography, to define all the device elements except the qubit. An Aluminum layer with a thickness of 200~nm is evaporated after resist developing and the step is finalized using lift-off. The second layer, which contains the qubit and the connections to the central line of the coplanar waveguide resonator, is realized using standard shadow evaporation of aluminum. The Josephson junctions are formed by two aluminum layers, with thickness 40~nm and 65~nm respectively, separated by an in-situ grown thin aluminum oxide layer.

The experiments are performed in a dilution refrigerator, at a temperature of 43~mK. The device is placed inside a copper box, connected to a printed circuit board by wire bonding. Connections to transmission lines are done using microwave launchers on the printed circuit board. Magnetic shielding is implemented using three layers of high magnetic permeability material. The sample is connected to room temperature electronics using coaxial cables, which include various filter, isolation, or amplification sections. The signal at the output port of the resonator is amplified using a low-noise high electron mobility transistor (HEMT) amplifier with a noise temperature of 4~K.

Readout and control pulses are implemented using modulation of continuous wave signals produced by synthesizers. Modulation signals are produced using arbitrary waveform generators with a time resolution of 1~ns and 4~ns for control/readout pulses respectively. The signal at the output of the resonator, amplified using the HEMT amplifier, is further amplified using an amplification chain at room temperature, demodulated, and digitized. The average of each readout output pulse is performed and recorded for each repetition. The time series of the digital measurement output are used to extract average quantities and the noise power spectral density.

Magnetic field biasing of the qubit is performed using a centimeter size coil attached to the copper box and fed by a current produced by a high stability current source.

%ACKNOWLEDGEMENTS

\emph{Author contributions:} MB and AL designed the experiment; MB fabricated the qubit/resonator device, conducted the experiment, and analyzed the data; JLO and FO contributed to the development of the experimental set-up; CD contributed to the development of software for data acquisition; AL and MB wrote the manuscript; all the authors discussed the data and commented on the manuscript.

\emph{Acknowledgements:} We are grateful to Mohammad Ansari, Jay Gambetta, Pieter de Groot, Seth Lloyd, Britton Plourde, and Frank Wilhelm for discussions. We acknowledge support from NSERC through Discovery and RTI grants, Canada Foundation for Innovation, Ontario Ministry of Research and Innovation, and Industry Canada. AL is supported by a Sloan Fellowship and JLO is supported by a WIN scholarship.

%REFERENCES
%\bibliography{magnetometerbib}%bibtex file
%\bibliographystyle{naturemag}
%\bibliography{D:/biblio/physics}%bibtex file

\end{document}